\pdfoutput=1
\documentclass{article}

\PassOptionsToPackage{numbers,sort&compress}{natbib}

\usepackage[preprint]{neurips_2026}

\usepackage[utf8]{inputenc}
\usepackage[T1]{fontenc}
\usepackage{amsmath,amssymb,amsfonts}
\usepackage{algorithm}
\usepackage{algpseudocode}
\usepackage{booktabs}
\usepackage{float}
\usepackage{graphicx}
\usepackage{hyperref}
\usepackage{microtype}
\usepackage{etoolbox}
\usepackage{tabularx}
\usepackage{tikz}
\usepackage{xcolor}
\usetikzlibrary{calc}
\hypersetup{hidelinks}

\graphicspath{{figures/}{figures/appendix/}}

\title{Local Consistency Does Not Guarantee Global Conservation:
Auditing Zero-Shot Composition of Airway Flow Operators}

\author{%
  \parbox{0.94\textwidth}{%
    \centering
    \small\bfseries
    Nichula Sathmith Wasalathilaka\textsuperscript{1},
    Navodya Heshan Samarasinghe\textsuperscript{1} \\
    Dhanujaya Suraweera\textsuperscript{2},
    Kevin Dawson\textsuperscript{2},
    Chinthaka Jacob\textsuperscript{2} \\
    Mervyn Parakrama Bandara Ekanayake\textsuperscript{1},
    Roshan Godaliyadda\textsuperscript{1}
    \\[0.35em]
    \normalfont\footnotesize
    \textsuperscript{1} Department of Electrical and Electronic Engineering,
    Faculty of Engineering, \\
    University of Peradeniya, Peradeniya, Sri Lanka \\
    \textsuperscript{2} Department of Mechanical Engineering,
    Faculty of Engineering, \\
    University of Peradeniya, Peradeniya, Sri Lanka
  }%
}

\makeatletter
\renewcommand{\@toptitlebar}{%
  \hrule height 4\p@
  \vskip 0.18in
  \vskip -\parskip%
}
\renewcommand{\@bottomtitlebar}{%
  \vskip 0.17in
  \vskip -\parskip
  \hrule height 1\p@
  \vskip 0.06in%
}
\patchcmd{\@maketitle}{\vskip 0.1in}{\vskip 0.05in}{}{}
\patchcmd{\@maketitle}
  {\vskip 0.3in \@minus 0.1in}
  {\vskip 0.14in \@minus 0.05in}{}{}
\renewcommand{\@notice}{\enlargethispage{2\baselineskip}}
\makeatother

\begin{document}

\maketitle

\begin{abstract}
  Neural operators approximate PDE solutions within a geometry family, but
independently learned local operators need not form a consistent global
simulator. We study frozen, single-pass composition for steady incompressible
flow in idealized two-dimensional airway trees. Separate Tube, bifurcation,
and trifurcation DeepONets are trained on 4,872 primitive CFD cases using field
supervision and auxiliary divergence, port-flux, component-balance, and
port-pressure penalties. The validation-selected deployment is frozen before
whole-tree CFD fields are inspected and assembled without tree training,
iterative coupling, flux correction, or CFD-informed adjustment. It retains
major flow patterns and controlled pathology responses with
\(0.204\)--\(0.215\)~s CPU inference, but has a \(22.68\%\)
prescribed-inlet-normalized external residual. A post-hoc sensitivity protocol,
frozen before new training and evaluation, repeats Data, Div, and Full models
across three seeds with Tube fixed. Relative to Data, Full reduces primitive
composite scores by \(26.7\%\) for Y2 and \(26.4\%\) for Y3 and reduces
assembled component-residual and interface-mismatch RMS by \(7.2\%\) and
\(16.6\%\).
Nevertheless, mean tree velocity error increases by \(7.5\%\), and external
residual increases from \(17.49\pm4.38\%\) to \(26.77\pm3.65\%\).
Local regularization can therefore improve primitive and assembled-local
diagnostics without ensuring accurate global fields or conservation.
\end{abstract}

\begin{figure}[H]
  \centering
  \vspace{-0.4em}
  \includegraphics[width=\linewidth]{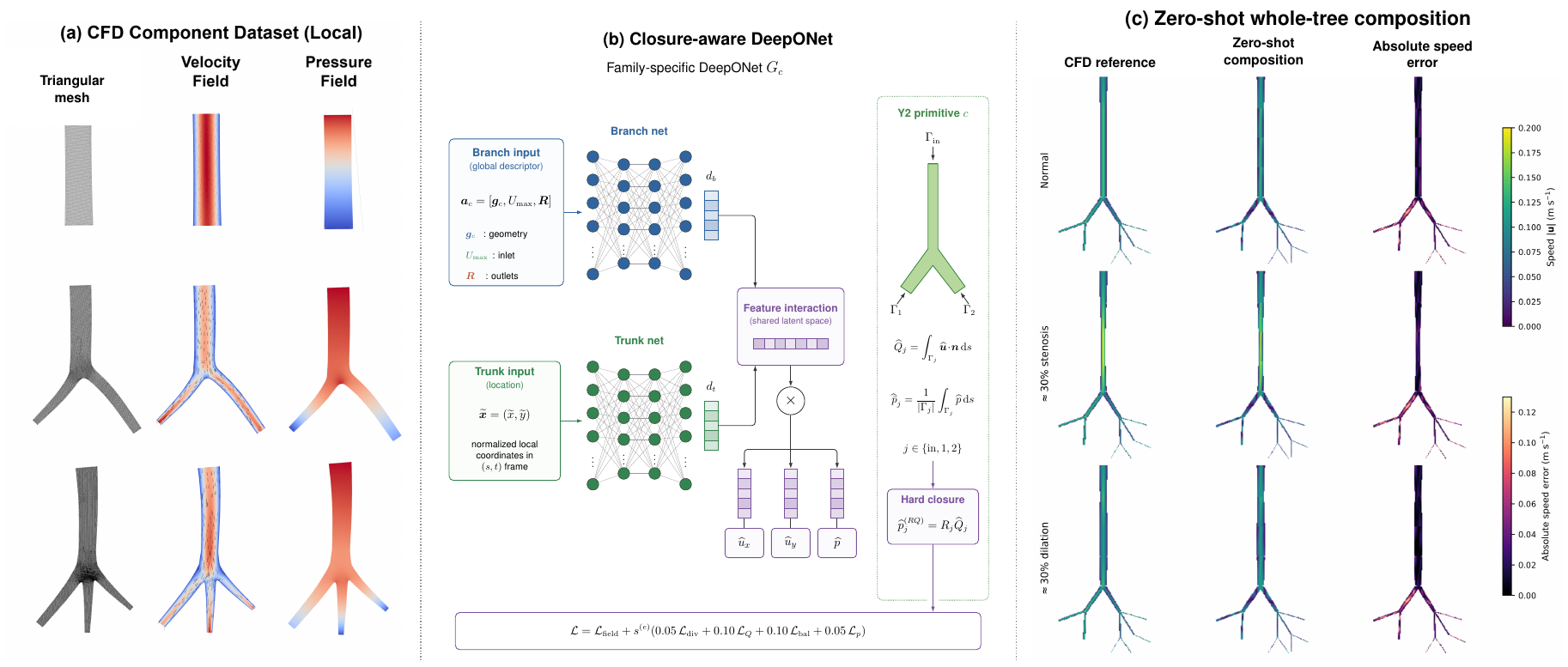}
  \caption{(a) Local primitive CFD. (b) Family-specific DeepONets are trained
  with primitive-level auxiliary constraints and then frozen. (c) Single-pass
  whole-tree deployment is evaluated against CFD; global conservation is
  audited, not enforced.}
  \label{fig:overview}
  \vspace{-0.5em}
\end{figure}

\section{Introduction}
\label{sec:introduction}

High-fidelity numerical simulation is essential for analysing flow and
transport in complex branching systems, but repeatedly solving geometrically
detailed domains can be computationally expensive. Learned surrogates amortize
this cost by mapping geometries and operating parameters to solution fields.
A full-domain surrogate, however, must cover the distribution of complete
configurations encountered at inference and offers limited reuse when similar
local structures recur across systems. This motivates our central question:
can operators learned independently on reusable physical primitives be frozen
and composed into an unseen larger system without full-domain retraining?

Neural operators provide a natural framework for this question. Deep Operator
Networks (DeepONets), Fourier Neural Operators (FNOs), and physics-informed
DeepONets learn mappings between input functions or parameterized physical
configurations and their solution fields
\citep{lu2021deeponet,li2021fno,
wang2021physicsinformeddeeponet}. Geometry-aware extensions such as GINO
accommodate varying irregular geometries, but still predict a solution over
each complete computational domain~\citep{li2023gino}. Generalization across
full-domain geometries is therefore distinct from independently learning local
component operators and assembling them into a previously unseen hierarchy:
the latter additionally requires predictions from different operators to be
mutually compatible at their interfaces.

Physics-informed objectives can improve local fidelity by incorporating
differential-equation residuals, boundary conditions, conservation penalties,
and interface or flux constraints
\citep{wang2021physicsinformeddeeponet,jagtap2020conservative}. These
constraints act within the domains on which they are imposed, whereas
composition introduces additional system-level requirements. Consequently,
low field and physics errors on isolated primitives do not imply a consistent
assembled solution:
\begin{equation}
  \text{local field accuracy}
  \;\not\Rightarrow\;
  \text{globally consistent composition}.
  \label{eq:local-global-gap}
\end{equation}
In an incompressible branching system, small component residuals can coexist
with parent--child flux mismatches, incompatible pressure gauges, or errors
that accumulate across successive generations.

Classical Schwarz methods obtain global consistency through iterative
transmission of interface information~\citep{gander2006optimizedschwarz}.
Recent learned local-to-global solvers retain explicit coupling: Schwarz
Neural Inference and Neural-Schwarz Tiling repeatedly update predictions over
overlapping subdomains~\citep{huang2026sni,secchi2026nest}, while CoMLSim
combines local representations with iterative inference
\citep{ranade2022composable}. M2PDE instead generates coupled multiphysics or
multi-component solutions by sampling from a joint generative model
\citep{zhang2025m2pde}. We study a deliberately stricter regime:
independently trained operators are frozen and evaluated once, then assembled
without iterative exchange, joint generation, flux correction, or global
projection. The contribution is therefore a controlled composition audit,
rather than a converged learned domain-decomposition solver.

We investigate this regime using steady two-dimensional airway flow. Airway
trees are organized as multigeneration tubular segments connected by branching
junctions~\citep{horsfield1971models,kitaoka1999airwaytree}. Following the
reusable primitive abstraction developed in RespGeomLib~\citep{11691379}, we
represent these structures as Tube, bifurcation (Y2), and trifurcation (Y3)
components, here idealized in two dimensions to permit controlled auditing of
component and interface errors. This setting preserves nontrivial velocity,
pressure, resistance, and interface physics while allowing component and
connection errors to be audited separately. The study is not a patient-specific respiratory model and does not
claim unrestricted topology generalization or prediction of physiological
terminal-flow redistribution. Here, \emph{zero-shot field composition} is
conditional on a predeclared terminal-flow allocation. Before any monolithic
solution field is accessed, each component inflow is obtained from the sum of
its descendant terminal flows and determines its parabolic inlet-speed scale;
resistance inputs follow the training-row matching rule documented in
Appendix~\ref{app:input-provenance}. No assembled tree enters training,
validation, selection, or correction, and Selected predictions are frozen
before monolithic CFD velocity and pressure fields are inspected. The post-hoc
sensitivity audit reuses primitive-selected checkpoints without tree-based
selection.

We contribute a controlled audit of frozen, single-pass operator composition.
The primary validation-selected deployment is archived before whole-tree CFD
fields are inspected. A subsequent sensitivity protocol, frozen before new
training and evaluation, repeats Data, Div, and Full branching operators over
three seeds while holding Tube fixed. The resulting audit separates primitive
consistency, assembled component and interface discrepancies, and the external
boundary budget. It shows that improvements in the first two levels do not
ensure improved global fields or conservation.
\section{Method}
\label{sec:method}

\paragraph{Primitive operators.}
An airway tree is represented as a directed graph
\(\mathcal T=(\mathcal V,\mathcal E)\), whose components are straight Tubes,
bifurcations (Y2), or trifurcations (Y3). For component family \(f_c\), an
independent DeepONet
\[
\mathcal G_{f_c}:
(\mathbf a_c,\widetilde{\mathbf x})
\mapsto
(\widehat{\mathbf u}_c,\widehat p_c)(\mathbf x)
\]
maps geometry, inlet-speed scale, outlet resistances, and normalized local
coordinate \(\widetilde{\mathbf x}\) to two-dimensional velocity and
pressure. The 4,872 geometry--operating-point cases comprise 1,000 Tube,
2,162 Y2, and 1,710 Y3 cases. Within each family, original-template groups
are disjoint across the 3,412/737/723 train/validation/test cases. Each case
stores 1,024 field samples; branch and target normalization statistics are
fitted on training cases only (Appendix~\ref{app:dataset}). No assembled tree
enters training, validation, selection, or refinement.

\paragraph{Closure-aware learning.}
For port \(k\in\mathcal P_c\), with cross-section \(\Gamma_{c,k}\) and
outward normal \(\mathbf n\), the signed flux and neural mean pressure are
\[
\widehat Q_{c,k}
=
\int_{\Gamma_{c,k}}
\widehat{\mathbf u}_c\!\cdot\!\mathbf n\,\mathrm ds,
\qquad
\widehat{\bar p}_{c,k}
=
\frac{1}{|\Gamma_{c,k}|}
\int_{\Gamma_{c,k}}\widehat p_c\,\mathrm ds .
\]
Inlet flux is negative, and
\(r_c=\sum_{k\in\mathcal P_c}\widehat Q_{c,k}\) is the component mass
residual. The Full model minimizes
\begin{equation}
\begin{aligned}
\mathcal L={}&\mathcal L_{\mathrm{field}}+s_\tau\!\left(
0.05\langle d^2\rangle+0.10\langle q^2\rangle+
0.10\langle b^2\rangle+0.05\mathcal L_p\right),\\
\ell^p_{c,k}&=
\frac{\widehat{\bar p}_{c,k}-\bar p^\star_{c,k}}{\sigma_{p,f_c}},
\qquad \mathcal L_p=\langle(\ell^p)^2\rangle .
\end{aligned}
\label{eq:closure-objective}
\end{equation}
The field term is equal-weight MSE over standardized \(u_x,u_y,p\); \(s_\tau\)
is zero through epoch 20 and ramps linearly to one at epoch 60.
\(\mathcal L_p\) compares the learned outlet-mean pressure with the stored CFD
target, uses no resistance, and excludes inlet pressure; only pilot Y2/Y3 cases
are supervised. Data uses field loss, Div adds divergence, and Full uses
Eq.~\eqref{eq:closure-objective}; primitive validation selects checkpoints or
fixed Data/Full blends. Validation, selection, and auditing additionally form
\(\widehat\pi^{RQ}_{c,k}=R_{c,k}\widehat Q_{c,k}\), which is not a pressure
output, does not replace \(\widehat p\), and does not enter gauge alignment
(Appendix~\ref{app:metrics}).

\paragraph{Frozen zero-shot composition.}
Each component is predicted in its canonical frame and placed in the tree by
\[
\mathbf x^{\mathrm g}
=
\mathbf o_c+\mathbf A_c\mathbf x^{\mathrm c},
\qquad
\widehat{\mathbf u}^{\,\mathrm g}_c
=
\mathbf A_c\widehat{\mathbf u}_c ,
\]
where \(\mathbf A_c\) and \(\mathbf o_c\) specify orientation and
translation. Pressure is not rotated. Algorithm~\ref{alg:compose-tree}
aligns only its additive gauge using raw neural port pressures and reports
conservation errors without modifying the predicted flow.

\begin{figure}[t]
\noindent
\begin{minipage}[t]{0.62\linewidth}
\vspace{0pt}
\footnotesize

\refstepcounter{algorithm}
\label{alg:compose-tree}
\noindent\textbf{Algorithm~\thealgorithm.}
\textsc{ComposeTree}: frozen zero-shot composition and audit
\vspace{-0.35em}

\algrenewcommand\algorithmicindent{0.85em}
\begin{algorithmic}[1]
\Require Tree \(\mathcal T=(\mathcal V,\mathcal E)\); component data
\(\{\mathbf a_c,\mathbf A_c,\mathbf o_c,f_c\}\); frozen
\(\{\mathcal G_f\}\)

\ForAll{\(c\in\mathcal V\)}
    \State \((\widehat{\mathbf u}_c,\widehat p_c)
    \gets\operatorname{Denorm}
    [\mathcal G_{f_c}(\mathbf a_c,\widetilde{\mathcal X}_c)]\)
    \State \((\{\widehat Q_{c,k},\widehat{\bar p}_{c,k}\}_k,r_c)
    \gets\operatorname{PortAudit}
    (\widehat{\mathbf u}_c,\widehat p_c)\)
\EndFor

\ForAll{\(e=(i,j)\in\mathcal E\)}
    \State \(m_e\gets
    \widehat Q_{i,e}^{\,\mathrm{out}}+
    \widehat Q_{j,e}^{\,\mathrm{in}}\)
\EndFor

\State \(\boldsymbol{\gamma}^{\star}\gets
\displaystyle\arg\min_{\gamma_{\mathrm{root}}=0}
\sum_{e=(i,j)\in\mathcal E}\Delta p_e(\boldsymbol{\gamma})^2\)
\Statex \hspace{\algorithmicindent}
\(\Delta p_e=
(\widehat{\bar p}_{i,e}^{\,\mathrm{out}}+\gamma_i)-
(\widehat{\bar p}_{j,e}^{\,\mathrm{in}}+\gamma_j)\)

\State \((\widehat{\mathbf u}^{\,\mathrm g},
\widehat p^{\,\mathrm g})
\gets\operatorname{Assemble}_{\mathrm{geom}}
\{\mathbf A_c\widehat{\mathbf u}_c,
\widehat p_c+\gamma_c^\star,\mathbf o_c\}_{c\in\mathcal V}\)

\State \(R_{\mathrm{ext}}\gets\sum_c r_c-\sum_e m_e\)
\State \(\varepsilon_{\mathrm{mass}}^{\mathrm{ref}}\gets
100|R_{\mathrm{ext}}|/Q_{\mathrm{in}}^{\mathrm{ref}}\)

\State \Return
\((\widehat{\mathbf u}^{\,\mathrm g},\widehat p^{\,\mathrm g})\),
\(\varepsilon_{\mathrm{mass}}^{\mathrm{ref}},\{r_c\},\{m_e\}\)
\end{algorithmic}
\end{minipage}
\hfill
\begin{minipage}[t]{0.37\linewidth}
\vspace{0pt}
\scriptsize
\raggedright

\refstepcounter{table}
\label{tab:implementation}
\noindent\textbf{Table~\thetable. Essential implementation details.}
\(D/F\): Data/Full; subscripts denote epochs.
\vspace{0.15em}

\setlength{\tabcolsep}{2pt}
\renewcommand{\arraystretch}{1.03}

\begin{tabularx}{\linewidth}{
  @{}
  >{\raggedright\arraybackslash}p{0.36\linewidth}
  >{\raggedright\arraybackslash}X
  @{}
}
\toprule
\textbf{Setting} & \textbf{Value} \\
\midrule

Operators
  & Independent Tube, Y2, and Y3 DeepONets \\

Network
  & Branch/trunk \(4{\times}256\), \(\tanh\); rank 128/field \\

Optimization
  & AdamW; LR \(10^{-3}\) (\(10^{-4}\) refinement);
    WD \(10^{-6}\); clip 1.0 \\

Batch / samples
  & 8 cases; 256 field and 64 divergence points/case \\

Schedule
  & 20-epoch warm-up; 40-epoch ramp; patience 60;
    primary seed 2026; sensitivity 2026--2028 \\

Loss weights
  & .05/.10/.10/.05 for divergence/flux/balance/port pressure \\

Port integration
  & 32-point midpoint on pilot caps; 32-point Gauss--Legendre at tree interfaces \\

Frozen selection
  & Tube refine (1); Y2 \(0.2D_{45}+0.8F_{85}\);
    Y3 \(0.3D_{45}+0.7F_{25}\) \\

\bottomrule
\end{tabularx}

\end{minipage}
\vspace{-0.4em}
\end{figure}

For interface \(e=(i,j)\), \(m_e\) is the parent--child flux mismatch.
The pressure offsets \(\gamma_c\) are determined only from neural interface
pressures, with \(\gamma_{\mathrm{root}}=0\); they change neither velocity
nor pressure gradients. Geometry-only ownership is used without spatial
blending, coordinate clipping, velocity rescaling, flux correction,
target-flow forcing, or CFD-informed adjustment.

The external residual
\(R_{\mathrm{ext}}=\sum_c r_c-\sum_e m_e\) and prescribed-inlet-normalized
imbalance
\(\varepsilon_{\mathrm{mass}}^{\mathrm{ref}}
=100|R_{\mathrm{ext}}|/Q_{\mathrm{in}}^{\mathrm{ref}}\), where
\(Q_{\mathrm{in}}^{\mathrm{ref}}
=3.74\times10^{-4}\,\mathrm{m^2\,s^{-1}}\), are diagnostic only.
Geometry, graph, prescribed boundary conditions, and component metadata are
predeclared; models, assembly rules, and predictions are frozen before
monolithic velocity and pressure CFD fields are inspected.
The three final monolithic-tree CFD references use
FEniCSx/DOLFINx~\citep{baratta2023dolfinx} with Taylor--Hood \(P_2/P_1\)
(vector-quadratic velocity/scalar-linear pressure) at bulk \(\mathrm{Re}=25.6\); the L40--L45
maximum pressure-drop change is below \(3.82\times10^{-4}\%\)
(Appendix~\ref{app:cfd}). Scoring uses one tree-wide pressure offset and
no component-wise CFD-fitted shifts.
\section{Results}
\label{sec:results}

\paragraph{Evaluation protocol.}
The Selected primitive checkpoints and tree arrays constitute the primary
prediction-before-CFD-field-inspection evaluation. The Data/Div/Full
comparisons are post-hoc after unblinding. Before training seeds 2027 and 2028,
we froze their configurations, seeds, checkpoint-selection rule, evaluation
order, and output locations. We then evaluated all 18 primitive checkpoints
and all 27 matched-seed tree deployments without result-based substitution.
Seed-2026 tree outputs reproduce the previous post-hoc arrays bitwise. All seed
summaries are computed from individual runs and report mean \(\pm\) sample SD;
the three controlled tree conditions are not treated as statistical replicates.

\paragraph{Primitive seed sensitivity.}
Across all three matched seeds, both Div and Full lower the composite test score
relative to Data for Y2 and Y3. Full lowers the mean score by \(26.7\%\) for Y2
and \(26.4\%\) for Y3, driven principally by divergence-MSE reductions of
\(90.5\%\) and \(75.0\%\). These improvements occur in all six matched
family-seed comparisons. Field and port changes are less uniform: mean velocity
changes are small, while Y3-Div increases flux RMSE by \(15.5\%\).
Auxiliary objectives therefore transfer robustly to the composite score and
divergence, but not uniformly to every metric.

\begin{figure}[t]
  \centering
  \includegraphics[width=0.68\linewidth]
  {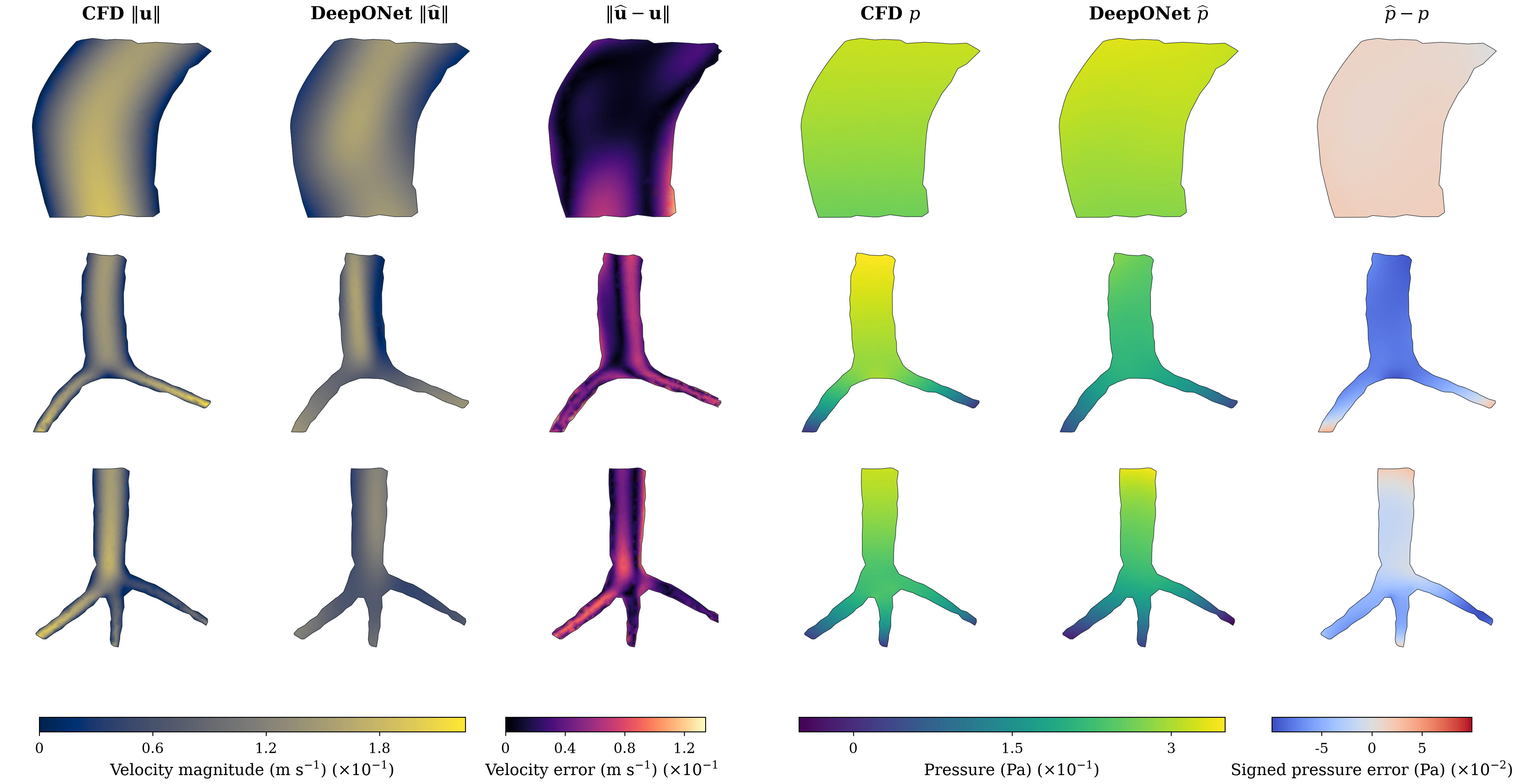}
  \caption{Held-out primitive predictions for Tube, Y2, and Y3:
  CFD and Selected DeepONet fields, velocity-vector error magnitude
  \(\lVert\widehat{\mathbf u}-\mathbf u\rVert_2\), and signed pressure
  error \(\widehat p-p\).}
  \label{fig:component-qualitative-validation}
\end{figure}

\begin{table}[t]
  \centering
  \caption{Primitive sensitivity. Data/Div/Full: mean \(\pm\) sample SD over
  seeds 2026--2028; \(\dagger\): original validation-fixed seed-2026 Selected
  point estimate. Lower is better.}
  \label{tab:primitive-test-audit}
  \scriptsize
  \setlength{\tabcolsep}{2.0pt}
  \renewcommand{\arraystretch}{0.97}
  \begin{tabular*}{\linewidth}{@{\extracolsep{\fill}}llccccc@{}}
    \toprule
    Family & Variant & Score & Vel. & Div. & Flux & Bal. \\
    \midrule
    Tube & Sel.\(\dagger\)
      & 0.269 & 0.337 & 0.078 & -- & -- \\
    \midrule
    Y2 & Data & \(0.437\pm0.024\) & \(0.406\pm0.006\)
      & \(1.926\pm0.326\) & \(0.158\pm0.028\) & \(0.302\pm0.039\) \\
       & Div & \(0.323\pm0.008\) & \(0.397\pm0.007\)
      & \(0.287\pm0.082\) & \(0.124\pm0.006\) & \(0.250\pm0.022\) \\
       & Full & \(0.320\pm0.009\) & \(0.403\pm0.014\)
      & \(0.183\pm0.054\) & \(0.127\pm0.014\) & \(0.181\pm0.044\) \\
       & Sel.\(\dagger\) & 0.310 & 0.381 & 0.171 & 0.126 & 0.232 \\
    \midrule
    Y3 & Data & \(0.417\pm0.027\) & \(0.434\pm0.024\)
      & \(2.352\pm0.195\) & \(0.149\pm0.012\) & \(0.252\pm0.065\) \\
       & Div & \(0.314\pm0.046\) & \(0.434\pm0.037\)
      & \(0.537\pm0.397\) & \(0.173\pm0.012\) & \(0.244\pm0.031\) \\
       & Full & \(0.307\pm0.046\) & \(0.420\pm0.041\)
      & \(0.589\pm0.274\) & \(0.127\pm0.037\) & \(0.178\pm0.051\) \\
       & Sel.\(\dagger\) & 0.325 & 0.421 & 1.075 & 0.139 & 0.211 \\
    \bottomrule
  \end{tabular*}

  \vspace{1pt}
  \parbox{\linewidth}{\scriptsize
  Score: common primitive audit; Vel.: vector Rel\(L_2\); Div.: dimensionless
  MSE; Flux: relative RMSE; Bal.: component RMS. Definitions:
  Appendix~\ref{app:metrics}; seed-level values and intervals:
  Appendix~\ref{app:detailed-results}.}
\end{table}

\paragraph{Blinded tree deployment.}
Figure~\ref{fig:overview}(c) and Table~\ref{tab:tree-results}(a) show that the
Selected composition retains axial-flow organization and stenosis
acceleration; CPU inference is \(0.204\)--\(0.215\)~s. Its external boundary
residual is \(22.676\%\), versus \(4.5\times10^{-7}\%\) for CFD. Prescribed
inlet normalization, alternate normalization, and pathology details are in
Appendices~\ref{app:signed-sums}, \ref{app:detailed-results},
and~\ref{app:branch-attribution}.

\begin{table}[t]
  \caption{Tree evaluation. (a) Selected deployment, frozen before CFD-field
  inspection. (b) Results averaged over the three controlled conditions
  within each seed, then reported as mean \(\pm\) sample SD across seeds
  2026--2028. Comp./Edge are
  \(100\,\operatorname{RMS}_{c}(r_c)/Q_{\mathrm{in}}^{\mathrm{ref}}\) and
  \(100\,\operatorname{RMS}_{e}(m_e)/Q_{\mathrm{in}}^{\mathrm{ref}}\) over all
  nine components and eight interfaces; External is condition-invariant.
  Conditions are not statistical replicates.}
  \label{tab:tree-results}
  \centering
  \scriptsize
  \renewcommand{\arraystretch}{0.95}

  \textbf{(a) Blinded Selected deployment}

  \setlength{\tabcolsep}{4.0pt}
  \begin{tabular*}{\linewidth}{
    @{\extracolsep{\fill}}lrrrrr@{}
  }
    \toprule
    Case &
    \(\mathbf u\) Rel\(L_2\) &
    Speed Rel\(L_2\) &
    \(p\) MAE &
    Peak speed &
    CPU \\
     & & & (Pa) & pred./CFD & (s) \\
    \midrule
    Normal
      & 0.594 & 0.376 & 0.085 & 0.135/0.157 & 0.209 \\
    \(\sim30\%\) stenosis
      & 0.554 & 0.348 & 0.094 & 0.181/0.197 & 0.204 \\
    \(\sim30\%\) dilation
      & 0.602 & 0.370 & 0.080 & 0.130/0.157 & 0.215 \\
    \bottomrule
  \end{tabular*}

  \vspace{0.1em}
  \textbf{(b) Three-seed post-hoc sensitivity}

  \setlength{\tabcolsep}{1.6pt}
  \begin{tabular*}{\linewidth}{@{\extracolsep{\fill}}lcccccc@{}}
    \toprule
    Variant & \(\mathbf u\) Rel\(L_2\) & Speed Rel\(L_2\) & \(p\) Rel\(L_2\)
    & Comp. RMS (\%) & Edge RMS (\%) & External (\%) \\
    \midrule
    Data & \(0.498\pm0.039\) & \(0.371\pm0.018\) & \(0.834\pm0.010\)
      & \(15.55\pm0.31\) & \(13.96\pm1.95\) & \(17.49\pm4.38\) \\
    Div & \(0.523\pm0.023\) & \(0.373\pm0.014\) & \(0.835\pm0.009\)
      & \(16.21\pm0.65\) & \(11.91\pm0.97\) & \(33.75\pm8.74\) \\
    Full & \(0.535\pm0.063\) & \(0.380\pm0.017\) & \(0.827\pm0.009\)
      & \(14.44\pm0.74\) & \(11.64\pm1.48\) & \(26.77\pm3.65\) \\
    \bottomrule
  \end{tabular*}
\end{table}

\paragraph{Three-seed tree sensitivity.}
Full lowers the case-averaged component-residual and interface-mismatch RMS
relative to Data in each of the three seeds; their mean reductions are
\(7.2\%\) and \(16.6\%\). Mean vector and speed Rel\(L_2\) increase by \(7.5\%\) and
\(2.4\%\), while pressure Rel\(L_2\) decreases by only \(0.9\%\).
External residual increases from
\(17.49\pm4.38\%\) for Data to \(26.77\pm3.65\%\) for Full and is higher
for Full in every seed. Div reaches \(33.75\pm8.74\%\).
Thus, improved primitive and assembled-local diagnostics do not guarantee
improved global fields or external conservation.

\paragraph{Pathology attribution.}
Within each variant-seed, the condition-invariant external residual measures
the external-port budget by telescoping and cannot localize internal error;
signed attribution is in Appendix~\ref{app:branch-attribution}.
\section{Limitations}
\label{sec:limitations}

This non-clinical study treats steady laminar flow in idealized 2D rigid-wall
airways, excluding patient anatomy, unsteadiness, wall motion, and turbulence.
One topology with prescribed inputs and terminal flows is tested, not arbitrary
topologies or physiological redistribution. Selected is one validation-fixed
seed-2026 blend. The three-seed Data/Div/Full protocol was frozen before new
training and evaluation but after original primitive-test and whole-tree CFD
unblinding; \(n=3\) gives descriptive, not inferential, variability. Pathology
conditions are not independent replicates. The external residual telescopes to
the external-port budget and cannot localize error; the assembler enforces
neither interface-flux continuity nor global balance. Unmatched hardware
precludes a normalized speedup claim.
\section{Conclusion}
\label{sec:conclusion}

Across three seeds, auxiliary objectives improve primitive composite scores and
divergence, while Full reduces assembled component-residual and
interface-mismatch RMS. Yet Full raises mean external residual from \(17.49\%\)
to \(26.77\%\); such gains do not ensure lower whole-tree field error or
imbalance. Frozen primitive operators are useful diagnostics, but reliable
composition requires conservative interface coupling, global projection, or
iterative correction rather than local regularization alone.

\clearpage
\bibliographystyle{plainnat}
\bibliography{references}

\clearpage
\appendix
\section{Additional material}
\label{app:additional-material}
\setcounter{figure}{0}
\renewcommand{\thefigure}{\thesection\arabic{figure}}
\renewcommand{\theHfigure}{appendix.\thesection.\arabic{figure}}

\paragraph{Broader impacts.}
Faster reusable flow surrogates could reduce the computational cost of
exploratory scientific simulation. However, the present model is not
clinically validated, and its pressure and conservation errors could lead to
unsafe conclusions if used for diagnosis, treatment planning, or device
design without independent CFD or experimental verification.

\subsection{Primitive dataset construction and splits}
\label{app:dataset}

The dataset contains 4,872 geometry--operating-point cases rather than 4,872
necessarily distinct geometries: multiple operating speeds can share one
original-template group. Family, source, split, and port-supervision counts are
given in Table~\ref{tab:primitive-dataset-splits}.

\begin{table}[ht]
  \centering
  \caption{Audited primitive dataset composition and leakage-safe splits.
  Entries are case or original-template-group counts.}
  \label{tab:primitive-dataset-splits}
  \scriptsize
  \setlength{\tabcolsep}{2.2pt}
  \renewcommand{\arraystretch}{1.03}
  \begin{tabular*}{\linewidth}{@{\extracolsep{\fill}}lrrrr@{}}
    \toprule
    Family &
    \shortstack{Legacy/\\pilot} &
    \shortstack{Train/val/test\\cases} &
    \shortstack{Train/val/test\\groups} &
    \shortstack{Port-supervised\\train/val/test} \\
    \midrule
    Tube  & 1000/0    & 700/148/152  & 175/37/38 & 0/0/0 \\
    Y2    & 1982/180  & 1523/326/313 & 351/76/73 & 132/24/24 \\
    Y3    & 1566/144  & 1189/263/258 & 280/61/59 & 96/24/24 \\
    Total & 4548/324  & 3412/737/723 & --        & 228/48/48 \\
    \bottomrule
  \end{tabular*}
\end{table}

The split-by-source case counts are Tube: 700/148/152 legacy; Y2:
1,391+132, 302+24, and 289+24 legacy+pilot for train, validation, and test;
and Y3: 1,093+96, 239+24, and 234+24. The grouping key is the original legacy
template identifier, and every group is assigned to exactly one split. When a
pilot case reuses a legacy template, the authoritative pilot assignment is
propagated to the matching legacy cases. This required 32 Y2 and 8 Y3 legacy
split-label overrides. Direct intersections of the train, validation, and test
group sets are empty for Tube, Y2, and Y3; cases are not independently split
point samples.

\paragraph{Field sampling and normalization.}
Every prepared case stores 1,024 field points. Legacy cases retain their frozen
prepared samples. Pilot points are sampled deterministically and uniformly with
respect to triangle area, followed by \(P_1\) interpolation of the CFD fields.
At each training epoch, 256 stored field points are randomly selected per case,
and 64 points are selected independently for divergence. Frozen primitive-test
evaluation uses all 1,024 field points and 256 divergence points per case.

Branch-feature normalization uses population means and standard deviations
fitted across training cases only. Target-channel population means and standard
deviations are fitted across all 1,024 spatial samples of the training cases
only. The same training-fitted statistics are applied without refitting to
validation and test cases.

\paragraph{Port supervision.}
Only pilot branch cases provide port supervision: Y2 has 180 supervised cases
split 132/24/24, Y3 has 144 split 96/24/24, and Tube has none. Legacy Y2/Y3
port losses are masked because exact port geometry is unavailable. Primitive
supervision uses 32-point fixed-order arclength midpoint quadrature on tagged,
mesh-realized CFD caps. Whole-tree inference instead uses 32-point
Gauss--Legendre quadrature on exact analytic interfaces; these are distinct
quadrature procedures.

\subsection{Loss, score, and metric definitions}
\label{app:metrics}

\paragraph{Standardized field loss and auxiliary objective.}
For output channel \(n\in\{u_x,u_y,p\}\), training uses
\[
\widetilde y_{ijn}=\frac{y_{ijn}-\mu_n}{\sigma_n},
\qquad
\mathcal L_{\mathrm{field}}
=\frac{1}{3BN_x}
\sum_{i=1}^{B}\sum_{j=1}^{N_x}\sum_{n=1}^{3}
\left(\widehat{\widetilde y}_{ijn}-\widetilde y_{ijn}\right)^2,
\]
so the three standardized channels receive equal pointwise weight. Here
\(N_x=256\) field points are sampled per training case; the 64 divergence
points are sampled independently. The auxiliary schedule is
\[
s_\tau=
\begin{cases}
0, & \tau\leq20,\\[1mm]
\min\!\left\{(\tau-20)/40,\,1\right\}, & \tau>20,
\end{cases}
\]
and therefore reaches full weight at epoch 60. The implemented Full objective
is
\[
\mathcal L_{\mathrm{Full}}
=\mathcal L_{\mathrm{field}}+s_\tau\left(
0.05\mathcal L_{\mathrm{div}}+
0.10\mathcal L_{\mathrm{flux}}+
0.10\mathcal L_{\mathrm{bal}}+
0.05\mathcal L_p\right).
\]
For case \(i\) and sampled point \(j\),
\[
d_{ij}=\frac{L_i}{U_i}\nabla\!\cdot\!\widehat{\mathbf u}_{ij},
\qquad L_i=\sqrt{W_iH_i},
\qquad
\mathcal L_{\mathrm{div}}=\operatorname{mean}_{i,j}d_{ij}^{\,2}.
\]
For outward-signed port fluxes,
\[
e^Q_{ik}=\frac{\widehat Q_{ik}-Q^\star_{ik}}
{|Q^\star_{i,\mathrm{in}}|},
\qquad
\mathcal L_{\mathrm{flux}}=\operatorname{mean}_{i,k}(e^Q_{ik})^2,
\]
and
\[
\mathcal L_{\mathrm{bal}}
=\operatorname{mean}_i\left(
\frac{\sum_k\widehat Q_{ik}}{|Q^\star_{i,\mathrm{in}}|}
\right)^2.
\]
The pressure term supervises the learned outlet pressure against the stored
primitive-CFD port target:
\[
\ell^p_{ik}=
\frac{\widehat{\bar p}_{ik}-\bar p^\star_{ik}}{\sigma_p},
\qquad
\mathcal L_p=\operatorname{mean}_{i,k\in\mathrm{out}}
\left(
\ell^p_{ik}
\right)^2.
\]
Here \(\widehat{\bar p}_{ik}\) is the quadrature-weighted mean of the learned
pressure field and \(\bar p^\star_{ik}\) is the stored CFD port-pressure
target. The training loss does not use \(R_{ik}\), does not form
\(R_{ik}\widehat Q_{ik}\), and does not supervise inlet pressure. Port losses
apply only to the pilot Y2/Y3 cases described in
Appendix~\ref{app:dataset}. Data uses only \(\mathcal L_{\mathrm{field}}\), Div
uses \(\mathcal L_{\mathrm{field}}+0.05s_\tau\mathcal L_{\mathrm{div}}\), and
Full uses all four auxiliary terms.

Training allowed at most 300 epochs, evaluated validation metrics every five
epochs, used a minimum learning rate of \(10^{-5}\), and stopped after 60
epochs without an improvement of at least \(10^{-5}\); the learning-rate
scheduler patience was six validation evaluations.

\paragraph{Common primitive audit score.}
The reported Score column is the post-training audit quantity
\[
S_f=
\overline{\mathcal E}_{\mathrm{field}}
+0.05\overline{\mathcal E}_{\mathrm{div}}
+\mathbb{I}_{f\in\{\mathrm{Y2},\mathrm{Y3}\}}
\left(
0.10\mathcal E_{\mathrm{flux}}+
0.10\mathcal E_{\mathrm{bal}}+
0.05\mathcal E_{RQ}
\right),
\]
where the field and divergence terms are evaluated on the frozen audit points.
For validation, selection, and diagnostic auditing, the evaluation-only
analytic decoder is
\[
\widehat\pi^{RQ}_{ik}=R_{ik}\widehat Q_{ik}.
\]
Its error diagnostic is
\[
\mathcal E_{RQ}
=\operatorname{mean}_{i,k\in\mathrm{out}}
\left(
\frac{\widehat\pi^{RQ}_{ik}-\bar p^\star_{ik}}{\sigma_p}
\right)^2.
\]
It is not a network pressure output, does not replace
\(\widehat p(\mathbf x)\), and is not used for tree pressure-gauge alignment.
For Tube, all three port
terms are absent. When both data sources exist, a bar denotes source-balanced
averaging,
\[
\overline z=\frac12\left[
\operatorname{mean}_{i\in\mathrm{legacy}}z_i+
\operatorname{mean}_{i\in\mathrm{pilot}}z_i
\right];
\]
Tube has only the legacy mean, and port metrics use only port-supervised pilot
cases. Training checkpoints used their variant-specific validation objectives,
whereas the reported primitive audit score applies one common post-training
definition; the two quantities should not be interpreted as identical.

\paragraph{Primitive metrics and uncertainty.}
For primitive case \(i\), the physical-unit velocity error is
\[
E_{\mathbf u,i}=
\left[
\frac{\sum_j\|\widehat{\mathbf u}_{ij}-\mathbf u_{ij}\|_2^2}
{\sum_j\|\mathbf u_{ij}\|_2^2}
\right]^{1/2},
\]
and the reported value is its source-balanced mean. Primitive pressure uses
training-set standardization without a fitted gauge shift,
\[
E_{p,i}^{\mathrm{std}}=
\left[
\frac1{N_x}\sum_j
\left(\frac{\widehat p_{ij}-p_{ij}}{\sigma_p}\right)^2
\right]^{1/2}.
\]
The reported pressure value is the source-balanced mean of
\(E_{p,i}^{\mathrm{std}}\).
The reported divergence is the source-balanced mean of per-case dimensionless
divergence MSE. The table reports
\(E_Q=\sqrt{\mathcal E_{\mathrm{flux}}}\),
\(E_{\mathrm{bal}}=\sqrt{\mathcal E_{\mathrm{bal}}}\), and the
\(RQ\)-decoder NRMSE
\(E_{RQ}=\sqrt{\mathcal E_{RQ}}\). Frozen testing uses 1,024 field and 256
divergence points per case.

Confidence intervals are percentile 95\% intervals from exactly 2,000
source-stratified, nonparametric geometry-group bootstrap replicates. Within
each source, geometry/template groups are sampled with replacement and all
cases in a selected group are retained. They quantify held-out group
variability for each fixed checkpoint, not training-seed uncertainty or paired
model-difference significance.

\paragraph{Whole-tree field metrics and pressure gauge.}
The frozen query grid has \(0.2\)-mm spacing and constant cell-area weights
\[
w_j=(0.2\times10^{-3}\,\mathrm m)^2.
\]
The velocity-vector and speed errors are
\[
E_{\mathbf u}=
\left[
\frac{\sum_jw_j\|\widehat{\mathbf u}_j-\mathbf u_j\|_2^2}
{\sum_jw_j\|\mathbf u_j\|_2^2}
\right]^{1/2},
\qquad
E_{|\mathbf u|}=
\left[
\frac{\sum_jw_j(\|\widehat{\mathbf u}_j\|_2-\|\mathbf u_j\|_2)^2}
{\sum_jw_j\|\mathbf u_j\|_2^2}
\right]^{1/2}.
\]
Exactly one scalar pressure offset is fitted over the complete tree,
\[
\delta=\frac{\sum_jw_j(p_j-\widehat p_j)}{\sum_jw_j},
\qquad
\bar p_w=\frac{\sum_jw_jp_j}{\sum_jw_j}.
\]
The reported gauge-adjusted pressure Rel\(L_2\) and MAE are
\[
E_p=
\left[
\frac{\sum_jw_j(\widehat p_j+\delta-p_j)^2}
{\sum_jw_j(p_j-\bar p_w)^2}
\right]^{1/2},
\qquad
\operatorname{MAE}_p=
\frac{\sum_jw_j|\widehat p_j+\delta-p_j|}{\sum_jw_j}.
\]
The same \(\delta\) applies to the entire tree; component-wise CFD-fitted
shifts are prohibited. Velocity is neither rescaled nor corrected, and the
evaluation-only \(RQ\)-decoder diagnostic is not used for pressure alignment.

\subsection{Whole-tree CFD reference simulations}
\label{app:cfd}

\paragraph{Scope and governing equations.}
This subsection documents only the three final L45 monolithic-tree CFD
references: normal, approximately 30\% stenosis, and approximately 30\%
dilation. They solve steady, two-dimensional, incompressible Newtonian
Navier--Stokes flow,
\begin{equation}
\rho(\mathbf u\cdot\nabla)\mathbf u
-\nabla\cdot\!\left(2\mu\boldsymbol{\epsilon}(\mathbf u)\right)
+\nabla p=\mathbf 0,
\qquad
\nabla\cdot\mathbf u=0,
\qquad
\boldsymbol{\epsilon}(\mathbf u)
=\tfrac12\left(\nabla\mathbf u+\nabla\mathbf u^\top\right).
\label{eq:tree-cfd-equations}
\end{equation}
The material parameters are
\(\rho=1.225~\mathrm{kg\,m^{-3}}\) and
\(\mu=1.7894\times10^{-5}~\mathrm{Pa\,s}\). The convective term is
retained, and no physical body force is applied.

\paragraph{Software, discretization, and licenses.}
The three solves used FEniCSx/DOLFINx 0.10.0.post2
\citep{baratta2023dolfinx}, Python 3.12.3, UFL 2025.2.1
\citep{alnaes2014ufl}, Basix 0.10.0, PETSc and petsc4py 3.19.6, mpi4py
3.1.5, and meshio 5.3.5 on one MPI rank. The Taylor--Hood discretization
uses vector continuous Lagrange \(P_2\) velocity and scalar continuous
Lagrange \(P_1\) pressure on first-order triangular geometry, without
stabilization. No quadrature degree was set manually; FEniCSx/FFCx defaults
were used. Gmsh generated the meshes~\citep{geuzaine2009gmsh}. The post-run
snapshot of the original CFD environment contained Gmsh 4.15.2; an
independently archived pre-run lock file proving its exact run-time version
was not retained.

The neural implementation used PyTorch 2.11.0+cu128
\citep{paszke2019pytorch}. Table~\ref{tab:software-licenses} records the
verified licenses of the major direct software assets. No external pretrained
model or external patient dataset is used.

\begin{table}[ht]
  \centering
  \caption{Recorded software versions, purposes, and verified licenses.}
  \label{tab:software-licenses}
  \scriptsize
  \setlength{\tabcolsep}{2.5pt}
  \renewcommand{\arraystretch}{1.02}
  \begin{tabularx}{\linewidth}{@{}l p{0.20\linewidth} X p{0.30\linewidth}@{}}
    \toprule
    Software & Recorded version & Purpose & License \\
    \midrule
    DOLFINx & 0.10.0.post2 & Whole-tree finite-element solve
      & LGPL-3.0-or-later \\
    UFL & 2025.2.1 & Weak-form specification
      & LGPL-3.0-or-later \\
    Gmsh & 4.15.2 post-run snapshot & Mesh generation
      & GPL-2.0-or-later with the Gmsh linking exception \\
    PyTorch & 2.11.0+cu128 & Neural training and inference
      & BSD-3-Clause \\
    PETSc/petsc4py & 3.19.6 & Linear algebra and solver interface
      & BSD-2-Clause \\
    Basix & 0.10.0 & Finite-element definitions
      & MIT \\
    \bottomrule
  \end{tabularx}
\end{table}

\paragraph{Boundary conditions.}
The inlet uses the two-dimensional parabolic profile
\[
\mathbf u(\mathbf x)
=-\mathbf n_{\mathrm{in}}U_{\max}
 \max(1-\eta^2,0),
\qquad
\eta=
\frac{(\mathbf x-\mathbf c_{\mathrm{in}})
      \cdot\mathbf t_{\mathrm{in}}}{R_{\mathrm{in}}}.
\]
Here \(U_{\max}=0.110~\mathrm{m\,s^{-1}}\),
\(R_{\mathrm{in}}=2.55~\mathrm{mm}\), and
\(D_{\mathrm{in}}=5.10~\mathrm{mm}\). Thus
\(U_{\mathrm{mean}}=(2/3)U_{\max}=0.0733333~\mathrm{m\,s^{-1}}\) and
\(Q_{\mathrm{in}}=U_{\mathrm{mean}}D_{\mathrm{in}}
=3.74\times10^{-4}~\mathrm{m^2\,s^{-1}}\). The reported bulk Reynolds
number is
\[
\mathrm{Re}
=\frac{\rho U_{\mathrm{mean}}D_{\mathrm{in}}}{\mu}
=25.60.
\]

All seven terminals use prescribed smooth parabolic outward-normal velocity
profiles, not pressure outlets or uniform profiles:
\[
\mathbf u_o
=\mathbf n_o U_{\max,o}\max(1-\eta_o^2,0),
\qquad
U_{\max,o}=\frac{1.5Q_o}{D_o}.
\]
The prescribed target flows are listed in
Table~\ref{tab:tree-cfd-terminal-flows}.

\begin{table}[ht]
  \centering
  \caption{Prescribed terminal target flows for the final whole-tree CFD
  references.}
  \label{tab:tree-cfd-terminal-flows}
  \small
  \setlength{\tabcolsep}{7pt}
  \begin{tabular}{lr}
    \toprule
    Terminal & \(Q_o~(\mathrm{m^2\,s^{-1}})\) \\
    \midrule
    \texttt{outlet\_L1} & \(1.17740740740741\times10^{-4}\) \\
    \texttt{outlet\_L2} & \(1.07351851851852\times10^{-4}\) \\
    \texttt{outlet\_R0} & \(4.1555555555556\times10^{-5}\) \\
    \texttt{outlet\_R1} & \(2.2509259259259\times10^{-5}\) \\
    \texttt{outlet\_R2} & \(3.2898148148148\times10^{-5}\) \\
    \texttt{outlet\_R3} & \(2.2509259259259\times10^{-5}\) \\
    \texttt{outlet\_R4} & \(2.9435185185185\times10^{-5}\) \\
    \bottomrule
  \end{tabular}
\end{table}

They satisfy
\[
\sum_o Q_o=3.74\times10^{-4}~\mathrm{m^2\,s^{-1}}
=Q_{\mathrm{in}}.
\]
All walls are no-slip. The parabolic opening profiles vanish at their
endpoints, so no separate corner correction was used. With
\(Q_\Gamma=\int_\Gamma\mathbf u\cdot\mathbf n\,\mathrm ds\), outlet flow is
positive and the raw inlet outward flux is negative.

\paragraph{Pressure reference.}
The CFD pressure null space was removed by fixing the \(P_1\) pressure degree
of freedom nearest the inlet centre to \(0~\mathrm{Pa}\). No zero-mean
pressure constraint or PETSc pressure-null-space object was used, and no
post-solution shift was applied to the saved CFD field. The saved CFD pressure
was not shifted during the solve. The single tree-wide scalar pressure
alignment used for error scoring is a separate evaluation operation; no
component-wise CFD-fitted shifts are permitted.

\paragraph{Nonlinear and linear solution.}
Picard fixed-point iteration started from \(\mathbf u^0=\mathbf 0\), without
continuation, ramping, or line search. It used at most 100 iterations,
under-relaxation \(\alpha=0.7\), and the criterion
\[
E_k=
\frac{\|\mathbf u^k-\mathbf u^{k-1}\|_2}
{\max(\|\mathbf u^k\|_2,10^{-14})},
\qquad
E_k<10^{-8}.
\]
The relaxed state update was
\[
\mathbf u_{\mathrm{prev}}
\leftarrow
0.7\mathbf u_{\mathrm{new}}
+0.3\mathbf u_{\mathrm{prev}}.
\]
The linear configuration was PETSc \texttt{preonly} with LU factorization,
requesting MUMPS when available and otherwise SuperLU\_DIST. The selected
factorization backend was not retained.

\paragraph{Mesh and mesh-independence study.}
Gmsh generated metre-based, first-order triangular meshes using
\texttt{Mesh.Algorithm = 6} and MSH 2.2 ASCII output, with no boundary-layer
elements and with local terminal refinement. L45 targets approximately 45
boundary elements across the smallest terminal diameter:
\[
D_{\min}=0.26~\mathrm{mm},
\qquad
h_{\mathrm{terminal}}=D_{\mathrm{terminal}}/45.
\]
The generated meshes contained approximately 47 boundary elements across the
smallest terminal. Their sizes are reported in
Table~\ref{tab:tree-cfd-meshes}.

\begin{table}[ht]
  \centering
  \caption{Final L45 whole-tree CFD meshes.}
  \label{tab:tree-cfd-meshes}
  \begin{tabular}{lrr}
    \toprule
    Case & Points & Triangles \\
    \midrule
    Normal & 178,313 & 344,886 \\
    Stenosis & 188,823 & 365,670 \\
    Dilation & 201,786 & 391,607 \\
    \bottomrule
  \end{tabular}
\end{table}

Mesh independence was assessed using the maximum relative L40-to-L45 change
among the seven inlet-to-outlet pressure drops, with a predefined threshold
below 1\%. The changes were \(0.0001882\%\) for normal, \(0.0003814\%\) for
stenosis, and \(0.0002722\%\) for dilation; this was not a comparison with an
analytical solution.

\paragraph{Numerical conservation and exported fields.}
Finite-element boundary integration gave final mass imbalances of
\(4.50655518\times10^{-7}\%\), \(4.50655503\times10^{-7}\%\), and
\(4.50655503\times10^{-7}\%\) for normal, stenosis, and dilation,
respectively, using
\[
100\,
\frac{\left|Q_{\mathrm{in}}-\sum_oQ_o\right|}
{|Q_{\mathrm{in}}|}.
\]
These numerical values come from finite-element boundary integration, not
merely the analytical target-flow sum. Velocity was solved in vector \(P_2\),
then interpolated to vector \(P_1\) for the exported nodal arrays. Pressure
was solved in scalar \(P_1\) and reordered to the Gmsh vertex ordering. Paper
field comparisons use these exported \(P_1\) nodal velocity and pressure
fields. Boundary fluxes were integrated from the original finite-element
velocity, not the exported \(P_1\) visualization field. Mean boundary pressure
was computed as
\[
\bar p_\Gamma=
\frac{\int_\Gamma p\,\mathrm ds}{\int_\Gamma\mathrm ds}.
\]

\paragraph{Timing and remaining provenance limits.}
The normal, stenosis, and dilation Picard-loop times were
\(498.52~\mathrm{s}\), \(512.42~\mathrm{s}\), and \(588.23~\mathrm{s}\),
respectively, on one MPI rank. Timing began immediately before the Picard loop
and stopped after the nonlinear solve; it excludes mesh generation, initial
mesh loading, complete boundary-condition setup, post-processing, and NPZ/VTU
writing. These unmatched-hardware timings are not compared directly with
neural CPU inference and do not support a hardware-normalized speedup claim.

The supplied records do not retain the CFD CPU model or core count, RAM or
peak memory, exact LU backend selected, exact pressure-gauge coordinates,
nonlinear iteration counts, total-script times, solver or mesh hashes, or exact
integrated per-port final flows.

\subsection{Whole-tree operating-input provenance}
\label{app:input-provenance}

Let \(\mathcal{T}(c)\) denote the prescribed terminal outlets downstream of
component \(c\). Its conditioning inflow is
\[
Q_c=\sum_{o\in\mathcal{T}(c)}Q_o,
\qquad
U_{\max,c}=\frac{3Q_c}{4r_c},
\]
where \(r_c\) is the exact operator-compatible inlet radius and the second
identity follows from the two-dimensional parabolic profile. These quantities
were constructed from boundary-condition metadata before any monolithic
solution field was opened and were not imposed on the predicted velocity.

Resistance inputs were neither obtained from whole-tree solution fields nor
computed by recursive resistance reduction. For each target component,
resistance entries were copied together from one training-split row selected
by nearest standardized geometry and inlet-speed features. Y2 and Y3 matching
additionally used an equally weighted normalized-conductance split-proxy block.
The \(1/R\) proxy was used only to select a plausible conditioning row and was
not treated as a terminal-flow predictor. The target geometry and
\(U_{\max,c}\) were retained exactly.

The input-support audit placed all target features within their training-split
marginal ranges except \(U_{\max,c}\) for the
\texttt{root\_tube\_2} and \texttt{root\_tube\_3} instances in the stenosis
condition. These are controlled marginal extrapolations; joint-distribution
support and arbitrary-topology generalization are not claimed.

\subsection{Detailed primitive and tree results}
\label{app:detailed-results}

The original Selected Tube, Y2, and Y3 operators were fixed using validation
data before primitive-test access, and their original Selected primitive
metrics were evaluated blind. Data, Div, and Full in the primitive audit were
evaluated post hoc, after primitive-test and monolithic-tree CFD solution-field
unblinding, using already frozen checkpoints. They did not alter checkpoints, blend weights,
selection, hyperparameters, or the original blind deployment. Reconstructing
the Selected Y2/Y3 blends reproduced every frozen aggregate metric and
2,000-replicate bootstrap interval exactly (maximum absolute difference
\(=0\)). All five protected source/result artifacts and both Div
checkpoints remained unchanged.

The tree audit likewise fixes the selected refined Tube operator in every row
and uses pre-existing primitive-validation checkpoints for Y2 and Y3. The
reconstructed Selected tree arrays are bitwise identical to the original blind
prediction arrays, and all original prediction and evaluation hashes remain
unchanged.

\paragraph{Three-seed sensitivity protocol.}
Before seeds 2027 and 2028 were trained or evaluated, the sensitivity registry
froze seeds 2026--2028, Data/Div/Full configurations, validation-only
checkpoint selection, matched Y2/Y3 deployment, evaluation order, and output
locations. For each seed and variant, Y2 and Y3 use matching
validation-selected checkpoints, while the selected Tube operator is fixed in
every tree run. All 18
primitive checkpoints and 27 tree deployments were evaluated without
test- or tree-result-based substitution. Seed-2026 tree predictions reproduce
the previous post-hoc Data/Div/Full arrays bit-for-bit.
Tables~\ref{tab:primitive-test-audit}--\ref{tab:tree-results} report
mean \(\pm\) sample standard deviation across seeds; controlled tree conditions
are first averaged within a seed and are not statistical replicates. The
detailed tables and figures below retain the original seed-2026 audit unless
explicitly identified otherwise.

\begin{table}[H]
  \caption{Seed-2026 primitive test audit. Score is the common post-training
  audit metric defined in Appendix~\ref{app:metrics}. Selected rows reproduce
  the original blind evaluation; Data, Div, and Full are secondary post-hoc
  diagnostics of frozen checkpoints. All metrics are lower-is-better.}
  \label{tab:supp-primitive-test-audit}
  \centering
  \scriptsize
  \setlength{\tabcolsep}{1.7pt}
  \renewcommand{\arraystretch}{0.98}
  \begin{tabular*}{\linewidth}{@{\extracolsep{\fill}}llrrrrrrr@{}}
    \toprule
    Family & Variant & Score & Velocity & Pressure & Divergence & Flux & Balance & \(RQ\) dec. \\
    \midrule
    Tube & Fixed Sel. & 0.269178 & 0.337450 & 0.083064 & 0.077469 & -- & -- & -- \\
    \midrule
    Y2 & Data     & 0.464965 & 0.408830 & 0.398393 & 1.860018 & 0.182180 & 0.346551 & 0.091558 \\
       & Div      & 0.328721 & 0.396934 & 0.332981 & 0.193215 & 0.131076 & 0.258782 & 0.067785 \\
       & Full     & 0.309975 & 0.387806 & 0.353782 & 0.121270 & 0.121437 & 0.213957 & 0.079152 \\
       & Selected & 0.310178 & 0.381449 & 0.346766 & 0.171014 & 0.125744 & 0.231625 & 0.076719 \\
    \midrule
    Y3 & Data     & 0.389745 & 0.418063 & 0.268420 & 2.224966 & 0.137017 & 0.176431 & 0.113189 \\
       & Div      & 0.367250 & 0.476292 & 0.279453 & 0.982131 & 0.186244 & 0.212262 & 0.111624 \\
       & Full     & 0.358716 & 0.467224 & 0.259273 & 0.899590 & 0.167268 & 0.236334 & 0.145284 \\
       & Selected & 0.324784 & 0.420967 & 0.239259 & 1.074812 & 0.138683 & 0.210766 & 0.125023 \\
    \bottomrule
  \end{tabular*}
\end{table}

\begin{figure}[!tbp]
  \centering
  \includegraphics[width=\linewidth]{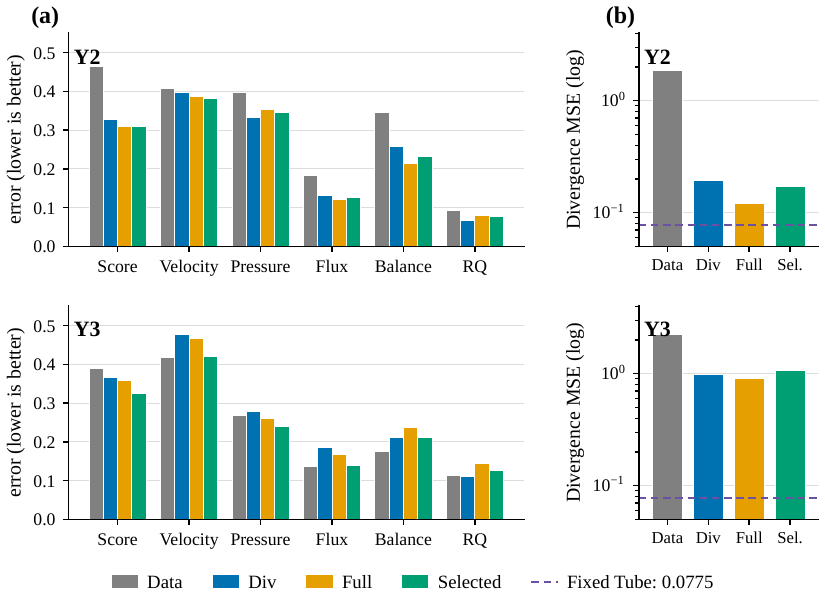}
  \caption{Seed-2026 primitive test audit for the frozen branching operators.
  Panel (a) compares primitive audit score, velocity, pressure, port-flux,
  component-balance, and evaluation-only \(RQ\)-decoder errors for Data, Div, Full, and
  Selected on Y2 and Y3; all metrics are lower-is-better. Panel (b) shows
  divergence MSE on a logarithmic scale, with the fixed Selected Tube
  divergence as a reference. Selected reproduces the blind evaluation;
  Data, Div, and Full are post-hoc diagnostics of frozen checkpoints. Exact
  values are reported in Table~\ref{tab:supp-primitive-test-audit}.}
  \label{fig:app-primitive-audit}
\end{figure}

Figure~\ref{fig:app-primitive-audit} provides the corresponding visual
comparison.

\begin{table}[H]
  \caption{Seed-2026 marginal 95\% intervals from 2,000 source-stratified
  template/group bootstrap replicates. They quantify held-out geometry-group
  variability for each fixed checkpoint, not training-seed uncertainty.
  Paired-difference intervals were not computed,
  so disjoint marginal intervals are not called statistically significant.
  Audit-score and divergence intervals are disjoint in both families; Y3 velocity
  intervals overlap.}
  \label{tab:supp-primitive-bootstrap}
  \centering
  \scriptsize
  \setlength{\tabcolsep}{4pt}
  \begin{tabular}{llrr}
    \toprule
    Family & Metric & Data 95\% interval & Selected 95\% interval \\
    \midrule
    Y2 & Audit score & [0.372584, 0.557496] & [0.270228, 0.349891] \\
       & Divergence & [1.747966, 1.980354] & [0.158439, 0.184643] \\
       & Velocity   & [0.397353, 0.420415] & [0.376390, 0.386641] \\
    \midrule
    Y3 & Audit score & [0.369387, 0.411504] & [0.290133, 0.360052] \\
       & Divergence & [2.034604, 2.418207] & [0.992298, 1.158008] \\
       & Velocity   & [0.398465, 0.438081] & [0.402319, 0.439646] \\
    \bottomrule
  \end{tabular}
\end{table}

\begin{figure}[!tbp]
  \centering
  \includegraphics[width=\linewidth]{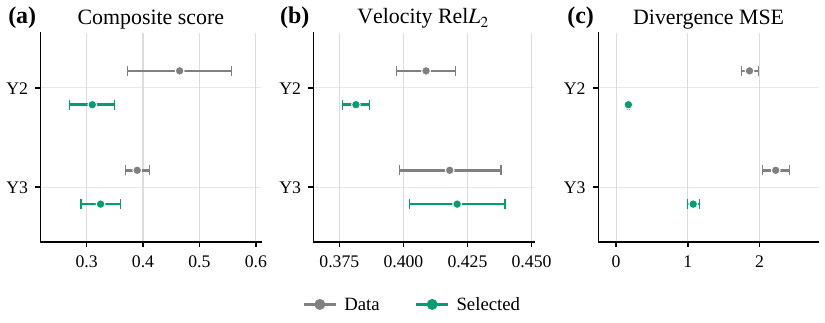}
  \caption{Marginal 95\% bootstrap intervals for Data and Selected on
  held-out Y2 and Y3 geometries. Intervals use 2,000 source-stratified
  template/group bootstrap replicates and quantify held-out geometry-group
  variability for each fixed checkpoint. They do not measure training-seed
  uncertainty and are not paired-difference confidence intervals. Lower
  values are better; exact endpoints are reported in
  Table~\ref{tab:supp-primitive-bootstrap}.}
  \label{fig:app-bootstrap}
\end{figure}

Figure~\ref{fig:app-bootstrap} visualizes these marginal intervals.

\begin{table}[H]
  \caption{Controlled pathology responses for the primary seed-2026 Selected
  deployment.}
  \label{tab:supp-pathology-response}
  \centering
  \small
  \setlength{\tabcolsep}{5pt}
  \begin{tabular}{lrr}
    \toprule
    Response & Prediction & CFD \\
    \midrule
    Stenosis \(\Delta u_{\max}\) (m/s)
      & \(+0.04508\) & \(+0.03986\) \\
    Stenosis \(\Delta p_{\mathrm{lesion}}\) (Pa)
      & \(+0.05550\) & \(+0.07433\) \\
    Dilation \(\Delta p_{\mathrm{lesion}}\) (Pa)
      & \(-0.02690\) & \(-0.02697\) \\
    \bottomrule
  \end{tabular}
\end{table}

\begin{figure}[!tbp]
  \centering
  \includegraphics[width=0.88\linewidth]{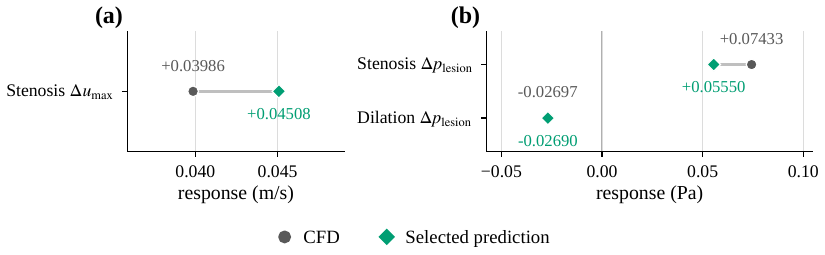}
  \caption{Controlled pathology responses of the frozen Selected deployment
  relative to the normal tree. Panel (a) shows the change in maximum velocity
  under approximately 30\% stenosis. Panel (b) shows lesion-pressure changes
  under stenosis and dilation. Gray circles denote CFD and green diamonds
  denote the Selected prediction; the dilation pressure-response markers are
  nearly coincident. Exact values are given in
  Table~\ref{tab:supp-pathology-response}.}
  \label{fig:app-pathology-response}
\end{figure}

The corresponding CFD and Selected response pairs are shown in
Figure~\ref{fig:app-pathology-response}.

\begin{table}[H]
  \caption{Complete seed-2026 per-case whole-tree field metrics.}
  \label{tab:supp-field-metrics}
  \centering
  \scriptsize
  \setlength{\tabcolsep}{4pt}
  \renewcommand{\arraystretch}{1.02}
  \begin{tabular*}{\linewidth}{@{\extracolsep{\fill}}llrrrr@{}}
    \toprule
    Model & Condition &
    \shortstack{Velocity\\Rel\(L_2\)} &
    \shortstack{Speed\\Rel\(L_2\)} &
    \shortstack{Pressure\\Rel\(L_2\)} &
    \shortstack{Pressure\\MAE (Pa)} \\
    \midrule
    Selected & Normal     & 0.59360 & 0.37563 & 0.82877 & 0.084993 \\
    Selected & Stenosis30 & 0.55423 & 0.34829 & 0.80073 & 0.094035 \\
    Selected & Dilation30 & 0.60231 & 0.36982 & 0.84015 & 0.079662 \\
    \addlinespace[0.15em]
    Data     & Normal     & 0.53377 & 0.36139 & 0.84677 & 0.087219 \\
    Data     & Stenosis30 & 0.49765 & 0.33473 & 0.81802 & 0.096390 \\
    Data     & Dilation30 & 0.53797 & 0.35405 & 0.85835 & 0.081642 \\
    \addlinespace[0.15em]
    Div      & Normal     & 0.55936 & 0.36752 & 0.83894 & 0.085989 \\
    Div      & Stenosis30 & 0.52187 & 0.34057 & 0.81049 & 0.095115 \\
    Div      & Dilation30 & 0.56555 & 0.36085 & 0.85042 & 0.080588 \\
    \addlinespace[0.15em]
    Full     & Normal     & 0.61839 & 0.37154 & 0.82326 & 0.084391 \\
    Full     & Stenosis30 & 0.57765 & 0.34439 & 0.79544 & 0.093395 \\
    Full     & Dilation30 & 0.62885 & 0.36531 & 0.83458 & 0.079119 \\
    \bottomrule
  \end{tabular*}
\end{table}

\begin{figure}[!tbp]
  \centering
  \includegraphics[width=0.95\linewidth]{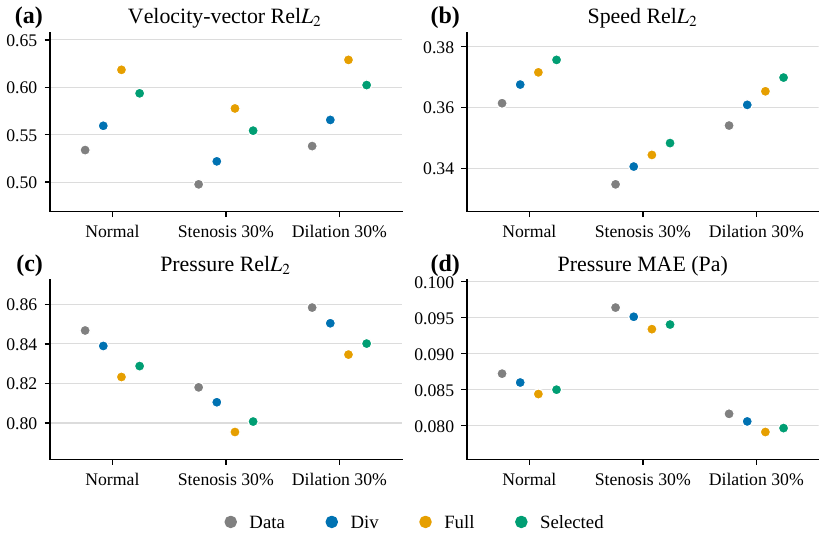}
  \caption{Per-condition whole-tree field errors for the frozen Data, Div,
  Full, and Selected branching variants with the same fixed Tube operator.
  Panels report (a) velocity-vector Rel\(L_2\), (b) speed Rel\(L_2\),
  (c) pressure Rel\(L_2\), and (d) pressure MAE. Values are descriptive
  results for the three controlled same-topology conditions (Normal,
  approximately 30\% stenosis, and approximately 30\% dilation), not
  independent replicate samples. Exact values are listed in
  Table~\ref{tab:supp-field-metrics}.}
  \label{fig:app-tree-fields}
\end{figure}

Figure~\ref{fig:app-tree-fields} shows the per-condition field metrics.

\subsection{Fixed-Tube and branching-only conservation attribution}
\label{app:branch-attribution}

This subsection is a secondary post-hoc attribution of the seed-2026 arrays;
the three-seed robustness result in Table~\ref{tab:tree-results} uses all nine components and all
eight interfaces.

The composed graph contains four Tube components followed by five Y2/Y3
components, with three Tube--Tube interfaces, one mixed Tube--Y2 interface,
four branch--branch interfaces, and eight external ports. Every audited
deployment uses the same protected Tube checkpoint; consequently, Tube residuals are
identical across Data, Div, Full, and Selected for a fixed condition.

For each deployment, the complete external-port flux vector is bitwise
identical across normal, stenosis, and dilation (maximum absolute difference
\(=0\); \texttt{array\_equal=True}). Branch-component RMS, branch--branch
interface RMS, and the external boundary residual are therefore each reported
once per deployment rather than averaged over the controlled conditions.
Field errors remain descriptive means because they vary by condition. The
single mixed Tube--Y2 edge is condition-dependent and is not averaged as a
pure branching-operator statistic.

With outward-signed port fluxes,
\[
r_c=\sum_{k\in\partial c}\widehat Q_{c,k},
\qquad
m_e=\widehat Q_{i,e}^{\mathrm{out}}+
\widehat Q_{j,e}^{\mathrm{in}},
\qquad
R_{\mathrm{ext}}
=\sum_c r_c-\sum_e m_e
=\sum_{b\in\partial\mathcal G}\widehat Q_b.
\]
Each internal port occurs once in the component sum and once in the interface
sum and thus cancels. Component RMS diagnoses component-level residual
magnitude, interface RMS diagnoses internal connection mismatch, and
\(R_{\mathrm{ext}}\) diagnoses only the external boundary-flux budget; it does
not localize internal error. Here it contains no pathology signal because the
external-port vectors do not change across conditions.

\begin{table}[H]
  \caption{Secondary seed-2026 conservation decomposition normalized by
  \(Q_{\mathrm{in}}^{\mathrm{ref}}\). Tube Comp. and Tube Edge contain the fixed
  Tube-chain contributions; Branch Comp. covers five Y2/Y3 components; Mixed
  Edge is the single Tube--Y2 interface; and Branch Edge covers four
  branch--branch interfaces. Branch-component, branch-edge, and external
  values are identical across the three conditions.}
  \label{tab:supp-group-decomposition}
  \centering
  \scriptsize
  \setlength{\tabcolsep}{2.0pt}
  \renewcommand{\arraystretch}{1.02}
  \begin{tabular*}{\linewidth}{@{\extracolsep{\fill}}llrrrrrr@{}}
    \toprule
    Model & Case &
    \shortstack{Tube\\Comp.} &
    \shortstack{Branch\\Comp.} &
    \shortstack{Tube\\Edge} &
    \shortstack{Mixed\\Edge} &
    \shortstack{Branch\\Edge} &
    External \\
    \midrule
    Data & Normal   & 13.608 & 10.830 & 10.720 & 31.214 & 11.201 & 21.266 \\
    Data & Stenosis & 22.567 & 10.830 & 16.381 & 40.063 & 11.201 & 21.266 \\
    Data & Dilation & 24.742 & 10.830 & 15.273 &  7.824 & 11.201 & 21.266 \\
    \addlinespace[0.15em]
    Div & Normal   & 13.608 & 10.640 & 10.720 &  8.379 &  6.147 & 39.685 \\
    Div & Stenosis & 22.567 & 10.640 & 16.381 & 17.229 &  6.147 & 39.685 \\
    Div & Dilation & 24.742 & 10.640 & 15.273 & 15.011 &  6.147 & 39.685 \\
    \addlinespace[0.15em]
    Full & Normal   & 13.608 & 2.227 & 10.720 &  6.178 & 3.618 & 23.147 \\
    Full & Stenosis & 22.567 & 2.227 & 16.381 & 15.028 & 3.618 & 23.147 \\
    Full & Dilation & 24.742 & 2.227 & 15.273 & 17.212 & 3.618 & 23.147 \\
    \addlinespace[0.15em]
    Selected & Normal   & 13.608 & 1.752 & 10.720 & 11.186 & 4.986 & 22.676 \\
    Selected & Stenosis & 22.567 & 1.752 & 16.381 & 20.035 & 4.986 & 22.676 \\
    Selected & Dilation & 24.742 & 1.752 & 15.273 & 12.205 & 4.986 & 22.676 \\
    \bottomrule
  \end{tabular*}
\end{table}

\begin{figure}[!tbp]
  \centering
  \includegraphics[width=\linewidth]{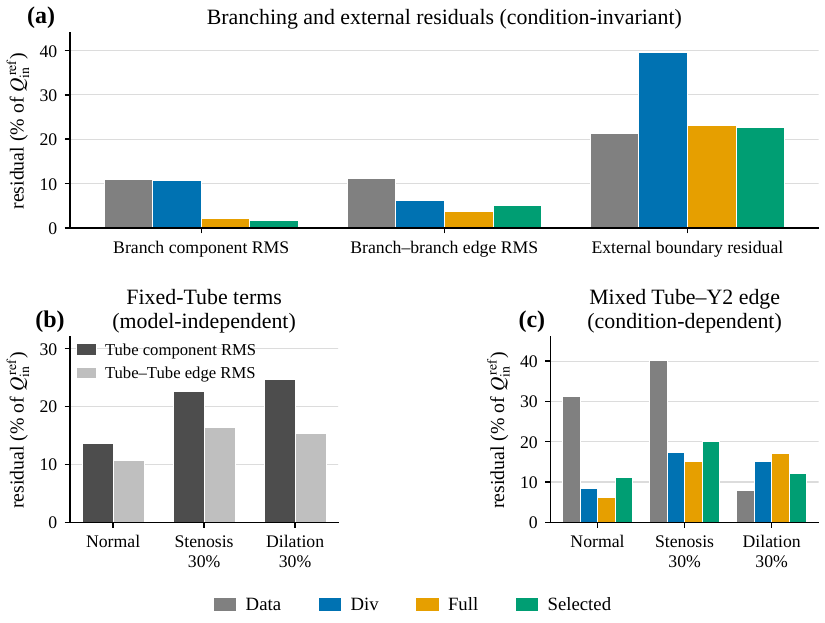}
  \caption{Conservation decomposition under the frozen tree audit,
  normalized by the prescribed inlet reference
  \(Q_{\mathrm{in}}^{\mathrm{ref}}\). Panel (a) shows branch-component RMS,
  branch--branch interface RMS, and external boundary residual for the four
  branching deployments; these quantities are condition-invariant for a
  fixed deployment. Panel (b) shows fixed-Tube component and Tube--Tube edge
  terms across the controlled conditions; these are model-independent because
  every deployment uses the same protected Tube operator. Panel (c) shows the
  single mixed Tube--Y2 interface, which is condition-dependent. Exact values
  are reported in Table~\ref{tab:supp-group-decomposition}.}
  \label{fig:app-conservation}
\end{figure}

The decomposition is visualized in
Figure~\ref{fig:app-conservation}.

\begin{table}[H]
  \caption{Secondary seed-2026 signed pathology-minus-normal changes as percentages of
  \(Q_{\mathrm{in}}^{\mathrm{ref}}\). Values are identical across Data, Div,
  Full, and Selected because the pathology modifies only instances of the
  fixed Tube operator. Equal component- and edge-sum shifts cancel from the
  telescoping external residual.}
  \label{tab:supp-pathology-attribution}
  \centering
  \small
  \setlength{\tabcolsep}{5pt}
  \begin{tabular}{lrr}
    \toprule
    Component/group & Stenosis--normal & Dilation--normal \\
    \midrule
    \texttt{root\_tube\_0} & 0.000 & 0.000 \\
    \texttt{root\_tube\_1} & \(+14.947\) & \(-18.034\) \\
    \texttt{root\_tube\_2} & \(+15.348\) & \(+26.293\) \\
    \texttt{root\_tube\_3} & \(-29.016\) & \(+28.707\) \\
    All Y2/Y3 components & 0.000 & 0.000 \\
    Component-sum shift & \(+1.280\) & \(+36.965\) \\
    Internal-edge-sum shift & \(+1.280\) & \(+36.965\) \\
    External-boundary shift & 0.000 & 0.000 \\
    \bottomrule
  \end{tabular}
\end{table}

The pathology-dependent signed shifts therefore arise entirely from three
instances of the shared fixed Tube operator, not from one component. For the
dilation case,
\(-18.033671+26.292509+28.706573=36.965410\%\); the equal internal-edge-sum
shift cancels it from the external boundary residual.

\begin{figure}[!tbp]
  \centering
  \includegraphics[width=0.90\linewidth]
    {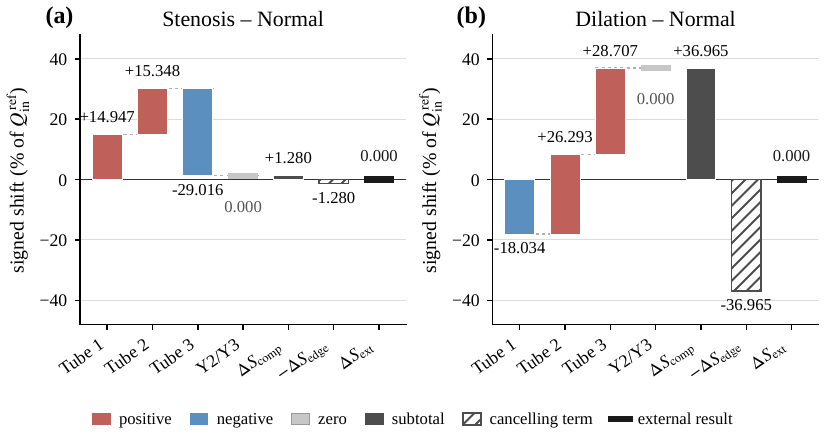}
  \caption{Signed pathology-minus-normal conservation shifts, expressed as
  percentages of \(Q_{\mathrm{in}}^{\mathrm{ref}}\). Panels (a) and (b) show
  stenosis--normal and dilation--normal, respectively. Pathology-dependent
  shifts arise from the shared fixed Tube instances, while the Y2/Y3 component
  contribution is unchanged. The component-sum shifts, \(+1.280\%\) and
  \(+36.965\%\), are matched by equal internal-edge-sum shifts, so
  \(\Delta S_{\mathrm{ext}}=0\) in both controlled conditions. The omitted
  root Tube has zero pathology-dependent change. Exact values are reported in
  Table~\ref{tab:supp-pathology-attribution}.}
  \label{fig:app-pathology-waterfall}
\end{figure}

Figure~\ref{fig:app-pathology-waterfall} shows this signed cancellation.

\subsection{Signed whole-tree sums and alternative normalization}
\label{app:signed-sums}

For each controlled tree condition, define
\[
S_{\mathrm{comp}}
=
100\frac{\sum_c r_c}{Q_{\mathrm{in}}^{\mathrm{ref}}},
\qquad
S_{\mathrm{edge}}
=
100\frac{\sum_e m_e}{Q_{\mathrm{in}}^{\mathrm{ref}}},
\qquad
S_{\mathrm{ext}}
=
S_{\mathrm{comp}}-S_{\mathrm{edge}}.
\]
RMS values in Table~\ref{tab:supp-conservation} quantify unsigned local
discrepancy magnitude, whereas the signed sums reconstruct the external
residual. The identity
\(S_{\mathrm{comp}}-S_{\mathrm{edge}}=S_{\mathrm{ext}}\) holds up to
numerical precision. Data's smaller external residual in this controlled
diagnostic does not imply uniformly smaller local errors. The
external/reference column uses the prescribed
\(Q_{\mathrm{in}}^{\mathrm{ref}}\); external/predicted-root is included only
as an additional normalization and is not the primary paper metric.

\begin{table}[H]
  \caption{Secondary seed-2026 whole-tree aggregate residual magnitudes, signed sums, and
  alternative normalizations. Aggregate RMS mixes four fixed Tube and five
  branching components and all eight internal interfaces; it is retained only
  for completeness. RMS values are in flow units; signed and external columns
  are percentages.}
  \label{tab:supp-conservation}
  \centering
  \scriptsize
  \setlength{\tabcolsep}{2.2pt}
  \renewcommand{\arraystretch}{1.02}
  \begin{tabular*}{\linewidth}{@{\extracolsep{\fill}}llrrrrrr@{}}
    \toprule
    Model & Case &
    \shortstack{Comp.\\RMS} &
    \shortstack{Edge\\RMS} &
    \(S_{\mathrm{comp}}\) &
    \(S_{\mathrm{edge}}\) &
    \shortstack{External/\\reference} &
    \shortstack{External/\\pred. root} \\
    \midrule
    Selected & Normal   & 3.428e-05 & 3.155e-05 & 15.829 & -6.847 & 22.676 & 30.602 \\
    Selected & Stenosis & 5.648e-05 & 4.778e-05 & 17.108 & -5.567 & 22.676 & 30.602 \\
    Selected & Dilation & 6.188e-05 & 4.072e-05 & 52.794 & 30.119 & 22.676 & 30.602 \\
    \addlinespace[0.15em]
    Data & Normal   & 4.542e-05 & 5.643e-05 & -21.180 & -42.446 & 21.266 & 28.699 \\
    Data & Stenosis & 6.385e-05 & 7.135e-05 & -19.900 & -41.166 & 21.266 & 28.699 \\
    Data & Dilation & 6.868e-05 & 4.699e-05 & 15.785 & -5.480 & 21.266 & 28.699 \\
    \addlinespace[0.15em]
    Div & Normal   & 4.507e-05 & 3.146e-05 & 60.084 & 20.399 & 39.685 & 53.556 \\
    Div & Stenosis & 6.360e-05 & 4.681e-05 & 61.363 & 21.678 & 39.685 & 53.556 \\
    Div & Dilation & 6.845e-05 & 4.338e-05 & 97.049 & 57.364 & 39.685 & 53.556 \\
    \addlinespace[0.15em]
    Full & Normal   & 3.449e-05 & 2.759e-05 & 26.413 & 3.266 & 23.147 & 31.238 \\
    Full & Stenosis & 5.661e-05 & 4.352e-05 & 27.692 & 4.545 & 23.147 & 31.238 \\
    Full & Dilation & 6.200e-05 & 4.281e-05 & 63.378 & 40.231 & 23.147 & 31.238 \\
    \bottomrule
  \end{tabular*}
\end{table}

\begin{figure}[H]
  \centering
  \includegraphics[width=0.92\linewidth]
    {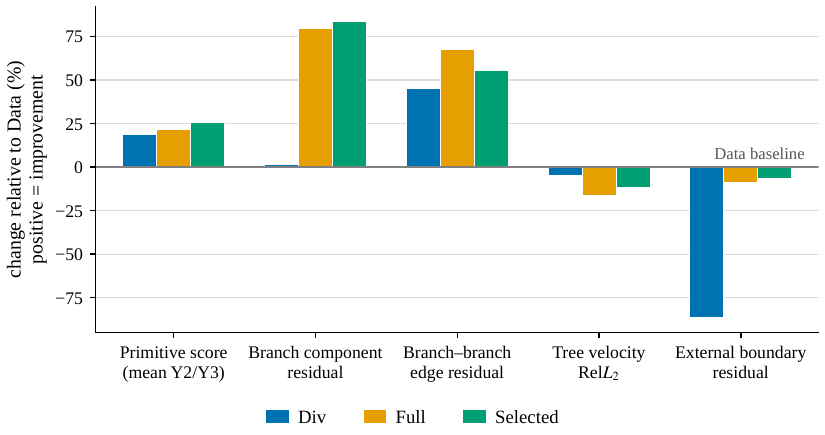}
  \caption{Secondary seed-2026 cross-scale change relative to the Data branching deployment. For
  each lower-is-better metric, the plotted quantity is
  \(100(\mathrm{Data}-\mathrm{model})/\mathrm{Data}\), so positive values
  denote improvement over Data and negative values denote degradation.
  Auxiliary objectives substantially improve primitive audit score and local
  branching conservation metrics, whereas the same ordering does not transfer
  to whole-tree velocity error or the external boundary residual. Data defines
  the zero baseline. This is a visualization of the frozen post-hoc audit and
  does not introduce an additional evaluation or model-selection criterion.}
  \label{fig:app-cross-scale}
\end{figure}

Figure~\ref{fig:app-cross-scale} summarizes how model ordering changes
between local and whole-tree metrics.

\paragraph{Code availability.}
Upon acceptance, we will release the complete training, evaluation,
composition, and figure-generation implementation in a public GitHub
repository, together with environment specifications and reproduction
instructions.

\end{document}